\def\HI{H{\,\small I}}
\newcommand{\msun}{{$M_\odot$}}
\newcommand{\msunyr}{{$M_\odot$ yr$^{-1}$}}

\documentclass[11pt,twoside]{article}
\usepackage{./asp2014}

\aspSuppressVolSlug
\resetcounters

\begin{document}

\title{Radio Jets Clearing the Way Through a Galaxy:
Watching Feedback in Action in the Seyfert galaxy IC~5063}
\author{R. Morganti$^{1,2}$, T.A. Oosterloo$^{1,2}$, J. B. R.  Oonk$^3$, W.
Frieswijk$^{1}$ and  C.N.Tadhunter$^{4}$}
\affil{$^1$ASTRON, the Netherlands Institute for Radio Astronomy, Postbus 2, 7990 AA, Dwingeloo, The 
Netherlands; \email{morganti@astron.nl}}
\affil{$^2$Kapteyn Astronomical Institute, University of Groningen, P.O. Box 800,
9700 AV Groningen, The Netherlands}
\affil{$^3$Leiden Observatory, Leiden University, P.O. Box 9513, 2300 RA Leiden, The Netherlands}
\affil{$^4$Department of Physics and Astronomy, University of Sheffield, Sheffield, S7 3RH, UK}

\paperauthor{Sample~Author1}{Author1Email@email.edu}{ORCID_Or_Blank}{Author1 Institution}{Author1 Department}{City}{State/Province}{Postal Code}{Country}
\paperauthor{Sample~Author2}{Author2Email@email.edu}{ORCID_Or_Blank}{Author2 Institution}{Author2 Department}{City}{State/Province}{Postal Code}{Country}
\paperauthor{Sample~Author3}{Author3Email@email.edu}{ORCID_Or_Blank}{Author3 Institution}{Author3 Department}{City}{State/Province}{Postal Code}{Country}

\begin{abstract}
High-resolution (0.5 arcsec) CO(2-1) observations performed with the
  Atacama Large Millimetre/submillimetre Array  have been used to trace the
  kinematics of the molecular gas in  the Seyfert 2 galaxy{IC~5063}. Although one of the most radio-loud Seyfert galaxy,  IC~5063 is a relatively
  weak radio source ($P_{\rm 1.4~GHz} = 3 \times 10^{23}$ W Hz$^{-1}$). The data
  reveal that the kinematics  of the gas is very complex. A fast
  outflow of molecular gas extends along the entire radio jet ($\sim 1$~kpc), with the highest outflow velocities about 0.5~kpc from the
  nucleus, at the location of the brighter hot-spot in the W lobe. All the observed characteristics can be described by a scenario of a radio
  plasma jet expanding into a clumpy medium, interacting directly with the
  clouds and inflating a cocoon that drives a lateral outflow into the
  interstellar medium. This suggests that most of the
  observed cold molecular outflow is due to fast cooling of the gas after the passage of
  a shock and that it is the end product of the cooling process.
  \end{abstract}

\section{Using gas outflows to trace the impact of AGN}

Understanding galaxy formation and evolution is one of the main challenges for
present day astronomy and is one of the key drivers for many of the new and future
large telescopes. The main paradigm for galaxy formation is that galaxies form by the
coalescence of smaller objects, and by the accretion of gas directly from its
environment \citep[see,
e.g.,][]{Bower2006}. One of the main complications is the tight
interplay between the star formation/AGN activity  and the surrounding ISM/IGM. 
This medium provides the material for forming stars and for fueling the activity of the
central black hole but, conversely, the enormous energy output from both the stellar
and the nuclear activity can have a major influence on the ISM/IGM and hence on the
further accretion of material onto the galaxy.
Observations are beginning to quantify the impact that the energy released by the AGN has on the host galaxy.
One of the processes playing a key role is gas outflows.
Particularly relevant is  the discovery that fast and massive outflows can also be traced by cold gas (\HI\ and CO, see e.g. \citet{Morganti2005,Feruglio2010,Dasyra2012,Morganti2013a,Cicone2014,Garcia2014}). This has challenged our ideas of how exactly the energy released by an AGN may interact with its surroundings. 

A number of different mechanisms have been suggested to originate the outflows \citep[see,
for overviews]{Fabian2013,Combes2014,Morganti2014}.
Among them, radio plasma jets are thought to be one of the main players in this complicated process.
Jets provide a particularly suitable way of transporting the energy.
They couple efficiently to the surrounding medium, injecting energy into the large scale
ISM/IGM,  therefore preventing gas to cool and form stars  \citep{McNamara2012}.
However, also on galaxy-scales they can have a major impact. Numerical simulations have shown how radio jets can  impact the ISM on scales that go well beyond the narrow beam of the jet at least in the case in which the plasma jet enters a clumpy medium \citep{Wagner2011,Wagner2012}.
What is new in this picture  is the presence of a component of cold (atomic and molecular gas) as tracer of this interaction and, even more interesting, the fact that this is the major component (in mass) of those outflows.
Thus, observations of cold molecular gas with ALMA have become a powerful way to quantify the characteristics and impact of this phase of the gas.

\articlefigure[width=13cm]{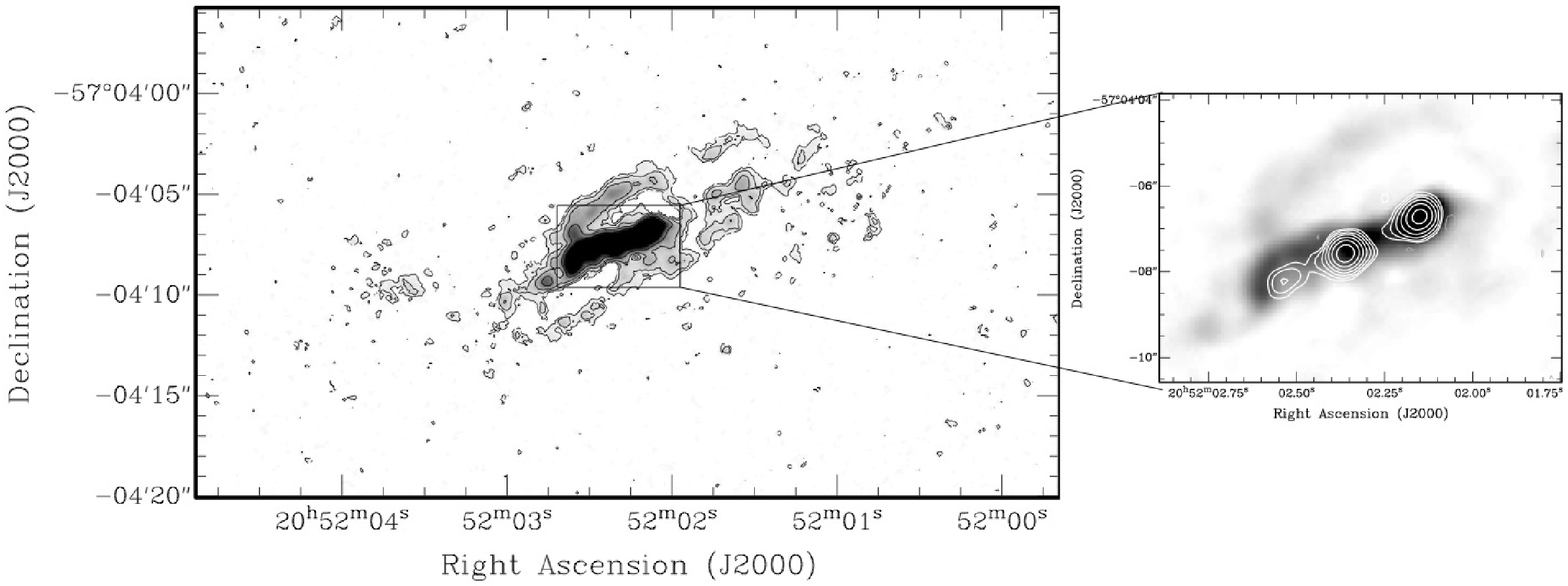}{fig1}{Total intensity of the CO(2-1) emission ( \emph{left}) and zoom in of the central region ( \emph{right}) with superimposed the contours of the 230~GHz  continuum emission.}

\section {IC~5063: our target and the ALMA observations}

We have imaged using ALMA the distribution and kinematics of the CO(2-1) emitting gas  in the nearby (z=0.011, 1$^{\prime\prime} \sim 200$ pc), southern radio-loud Seyfert galaxy
IC5063. This was the first object where a fast outflow of neutral hydrogen was discovered (Morganti et al. 1998).
Follow up VLBI observations allowed us to establish that the region where the outflow is
occurring coincides with the bright western radio lobe, roughly 0.5~kpc from the nucleus (Oosterloo et
al. 2000).  Extremely exciting has been  the finding (using ISAAC on the VLT) that this fast  outflow also contains warm molecular gas (H$_2$ at 2.2$\mu$m) co-spatial with the bright radio hot-spot \citep{Tadhunter2014}. This reinforces the idea of a prominent role of the radio jet as driver of these outflows.  APEX revealed that also cold molecular gas is associated with the outflow \citep{Morganti2013b} but, given the poor spatial resolution, could not locate the gas. This is the goal of the ALMA observations.

IC~5063 was observed during Cycle 1 with ALMA using Band 6, covering simultaneously the CO(2-1) centered
on 227.968~GHz and two additional base-bands (at 232 and 247~GHz) for imaging of the continuum.
In Fig.\ref{fig1} the total intensity of the CO(2-1) is shown with a zoom-in of the central region and overlaid the mm continuum image. The spatial resolution obtained is $\sim 0.5$ arcsec and the r.m.s. noise per channel in the CO cube is  0.3 mJy beam$^{-1}$.

\section{Signatures of a complex interaction}

The  ALMA observations give a detailed view of the complexity of the distribution and kinematics of the molecular gas in the
central regions (see Fig. \ref{fig2}). While the large-scale structure follows a regularly rotating  disk, {\sl the gas in the inner regions is strongly affected by the radio jet}. Indeed, the continuum emission is clearly co-spatial with the brighter inner-region detected in CO (see Fig.\ref{fig1}, right).
Thus, the ALMA data confirm the multi-phase of the gaseous outflow in IC~5063, adding the cold molecular gas to the \HI, warm molecular and ionized gas  (see \citet{Tadhunter2014} and refs. therein).  

\articlefigure[width=6.5cm]{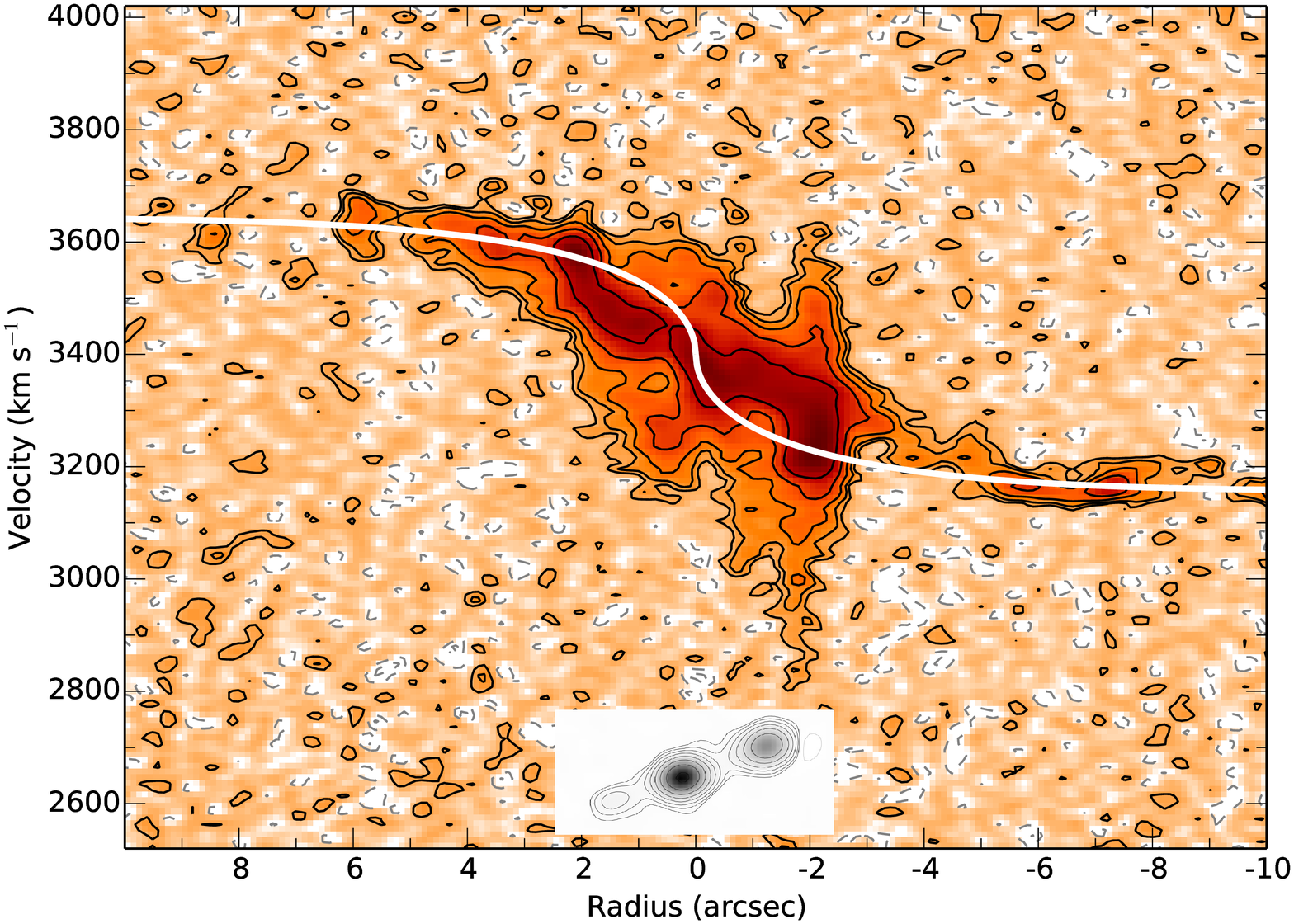}{fig2}{Integrated (over 2$^{\prime\prime}$ perpendicular to the major axis) position-velocity map taken along the major axis of IC~5063. The white line gives the rotation curve we derived from the photometry of \citet{Kulkarni1998} and  illustrates the expected kinematics of gas following regular rotation. The  deviation from this in the inner region (i.e. region co-spatial with the continuum emission) is evident.}

The observed distribution and kinematics of the cold molecular gas suggest that  the radio plasma jet is driving the bulk of the gas outflow.   Therefore, we have developed a simple model to describe the kinematics of the gas, following the scenario presented in the numerical simulations of \citet{Wagner2011} and \citet{Wagner2012}. These simulations describe the effects of a newly formed radio jet when moving though a {\sl dense clumpy medium}. This medium forces the jet to find the path of least resistance, while interacting and gradually dispersing the dense clouds away from the jet axis.  In this way, clouds can be accelerated to high
velocities and over a wide range of directions. Along the path of the jet, a turbulent cocoon of expanding gas forms, moving away from the jet axis. 
In our toy-model, we introduce a lateral expansion of the gas, away from the jet axis. This also causes the rotation velocity of the affected gas to be well below the nominal rotation of the gas in the disk, as indeed observed (see Fig. 2).  Despite its simplicity, the model provides a good first-order description of the observations.
The full description of the results and of the model can be found in Morganti et al. (submitted). We argue that the origin of the cold gas would be related  to the efficient cooling of the gas after the shock produced by the jet: cold molecular gas would represent the final product of this cooling process, while warm molecular and \HI\ would be  intermediate (and less massive) phases.  

Even assuming the most conservative values for the conversion factor CO-to-H$_2$, we find that the mass of the outflowing gas is between $1.9$ and $4.8 \times 10^7$ \msun, of which between $0.5$ and $1.3 \times 10^7$ \msun\ is associated  with the fast outflow at the location of the W lobe, i.e. much larger that what found in the other phases of the gas.  Using these values, we
derive a mass outflow rate  ($\dot{M} = M/\tau_{\rm dyn}$) in the range 12 to 30 \msunyr. Interestingly, both the radiation as well as the jet power would be  energetically capable of driving the outflow. The fact that the radio jet plays the prominent role may suggest an higher efficiency of this mechanism or that this mechanism is more relevant at large distances from the nucleus. 

Particularly relevant is the conclusion that the effect of the radio plasma can be significant also in objects, such as IC~5063, that are often considered {\sl radio-quiet}. A number of cases where this situation may occur have been recently identified (e.g.\ NGC~1266 \citet{Alatalo2011}; NGC~1433 \citet{Combes2013} and M51 \citet{Matsushita2015}), although IC~5063 represents the most clear case of such an interaction being on-going. Because  weak radio sources are more  common than the powerful ones, this result has implications for the role of feedback in galaxy evolution.

\acknowledgements This paper makes use of the following
ALMA data: \\
ADS/JAO.ALMA\#2012.1.00435.S. ALMA is a partnership
of ESO (representing its member states), NSF (USA) and NINS (Japan), together
with NRC (Canada) and NSC and ASIAA (Taiwan), in cooperation with the Republic
of Chile. The Joint ALMA Observatory is operated by ESO, AUI/NRAO
and NAOJ. RM gratefully acknowledge support from the European Research Council under the European Union's Seventh Framework Programme (FP/2007-2013) /ERC Advanced Grant RADIOLIFE-320745.



\end{document}